\documentclass[letterpaper, 10 pt, conference]{ieeeconf}  

\IEEEoverridecommandlockouts                              
\usepackage[utf8]{inputenc}
\usepackage{amsmath, amssymb}
\usepackage{amsfonts}
\usepackage{mathtools}
\usepackage{enumerate}
\usepackage{tikz}
\usepackage{subfigure}
\usepackage{graphics}
\usepackage[noend]{algpseudocode}
\usepackage{algorithm}
\usepackage{algorithmicx}
\newcommand*\Let[2]{\State #1 $\gets$ #2}
\algrenewcommand\algorithmicrequire{\textbf{Input:}}
\algrenewcommand\algorithmicensure{\textbf{Output:}}

\newcommand{\R}{\mathbb{R}}

\def\balign{\begin{align*}}
\def\ealign{\end{align*}}
\def\bbmat{\begin{bmatrix}}
\def\ebmat{\end{bmatrix}}
\newcommand{\cm}{\exp(-\Delta t \mu)}

\newtheorem{theorem}{Theorem}
\newtheorem{example}{Example}
\newtheorem{lemma}{Lemma}
\newtheorem{corollary}{Corollary}

\newtheorem{proposition}{Proposition}
\newtheorem{assumption}{Assumption}
\newtheorem{definition}{Definition}

\newtheorem{remark}{Remark}
\def\BT{\begin{theorem}}
\def\ET{\end{theorem}}
\def\BL{\begin{lemma}}
\def\EL{\end{lemma}}
\def\BP{\begin{proposition}}
\def\EP{\end{proposition}}
\def\BC{\begin{corollary}}
\def\EC{\end{corollary}}
\def\BD{\begin{definition}}
\def\ED{\end{definition}}
\def\BA{\begin{assumption}}
\def\EA{\end{assumption}}
\def\BR{\begin{remark}}
\def\ER{\end{remark}}
\def\BE{\begin{example}}
\def\EE{\end{example}}

\title{\LARGE \bf
Mean Reverting Portfolios via Penalized OU-Likelihood Estimation 
}

\author{Jize Zhang, Tim Leung and Aleksandr Aravkin
\thanks{Jize Zhang, Tim Leung and Aleksandr Aravkin are with the Department of Applied Mathematics, University of Washington, Seattle WA.}%
\thanks{A. Aravkin's work is supported by the WRF Data Science Professorship.}
}

\begin{document}

\maketitle
\thispagestyle{empty}
\pagestyle{empty}

\begin{abstract}

We study  an optimization-based approach to construct a mean-reverting portfolio of assets. 
Our objectives are threefold: (1) design a portfolio that is well-represented by an Ornstein-Uhlenbeck process 
with parameters estimated by maximum likelihood, (2) select portfolios with desirable characteristics of high mean reversion and low variance, 
and (3) select a parsimonious portfolio, i.e. find a small subset of a larger universe of assets that can be used for long and short positions. 
We present the full problem formulation, a specialized algorithm that exploits partial minimization, 
and numerical examples using both simulated and empirical price data.


\end{abstract}

\section{Introduction}
A major class of trading strategies in various financial  markets, including equities and commodities, is based on taking advantage of the mean-reverting behavior of asset prices. In practice, a portfolio of simultaneous positions in two or more highly correlated or cointegrated assets, such as stocks and exchange-traded funds (ETFs),  can be constructed to obtain mean-reverting prices. There are a number of studies on the empirical performance of pairs trading~\cite{gatev2006pairs} and timing of trades~\cite{elliott2005pairs,meanreversionbook2016,LeungLi2014OU,KitaLeung}. 
There are also approaches for identifying mean-reverting portfolios with a few assets from a larger collection of stocks \cite{d2011identifying}. 

Given a set of assets and time series of their historical prices,  our main goal is to design a mean-reverting portfolio whose evolution over time can be characterized by an Ornstein-Uhlenbeck (OU) process~\cite{Ornstein1930} through penalized likelihood estimation. The OU process is continuous-time version of the discrete-time autoregressive (AR) model \cite{engle1987co}.  

A major feature of our joint optimization approach is that we \emph{simultaneously} solve for  the optimal portfolio and the corresponding parameters for maximum likelihood. This unified approach is different from prior work for OU portfolio selection, which break the problem up into stage-wise computations. For example,~\cite{d2011identifying} first finds an OU representation for the time series of multiple assets, 
and then solves a second optimization problem to find the portfolio based on that representation. Conversely,~\cite{meanreversionbook2016} fits an OU process to each of a range of candidate (pair) portfolios, and takes the candidate with the highest OU likelihood.  Our unified approach looks for the best OU-representable portfolio 
from a set of candidates,  making the quality of the OU fit part of the optimization problem. 
The OU characterization of the discovered portfolio then
informs the optimal trading strategies, such as those developed in~\cite{meanreversionbook2016}.

The paper proceeds as follows. In Section~\ref{sec:model} we derive the same maximum likelihood formulation from two viewpoints: a continuous stochastic differential equation (SDE) characterization, 
and a discrete-time autoregressive (AR) approximation. 
 We then modify the MLE formulation to include terms that promote portfolio sparsity and high mean reversion, as well as 
 terms that select fewer assets from a larger candidate set.
 In Section~\ref{sec:algo} we develop an algorithm for the nonsmooth, nonconvex objective 
based on partial minimization and projection, and show that it performs much better than a 
standard algorithm that doesn't exploit problem structure. 
In Section~\ref{sec:numerics} we provide numerical illustrations using 
 both simulated and real data. We end with a discussion in Section~\ref{sec:discussion}.

\section{Problem Formulation }
\label{sec:model}

In this section, we derive the maximum likelihood formulation for simultaneously selecting a portfolio 
from a set of assets, and representing that selection using an Ornstein-Uhlenbeck (OU) process. 
We first show that the same likelihood formulation is obtained considering either the continuous SDE characterization 
of the OU, and of its discretized autoregressive (AR) approximation, with a subtle difference in parameter interpretation. 
We also make several theoretical observations 
about the well-posedness of the estimation problem. We then extend the maximum likelihood formulation 
to allow selection of lower variance, higher mean reversion, and parsimony in the portfolio. 

\subsection{MLE Derivations}

We are given historical data for $m$ assets, with $S^{(T+1)\times m}$ the matrix for assets values over time. 
Our main goal is to find $w$, the linear combination of assets that comprise our portfolio, such that  the corresponding portfolio  price process $x_t := S_tw$ best follows an OU process. We first show that solving for the portfolio 
and OU likelihood, whether  using the SDE characterization of OU or the AR characterization, 
both yield the joint objective 
\begin{equation} 
\label{eq:mlegen}
\min_{a, c, \theta,\|w\|_1 = 1} \frac{1}{2} \ln(a)  + \frac{1}{2T a} \| A(c)w - \theta(1- c) \|^2,
  \end{equation}
where $A(c) = S_{1:T} - cS_{0:T-1}$, $w$ is the portfolio to be selected,  and $a, c, \theta, \sigma$ are likelihood parameters. 
The objective function is nonconvex, since $A(c)$ multiplies $w$, and also 
includes a nonconvex constraint $\|w\|_1 = 1$.  The derivations are presented below.

\noindent
{\bf Maximum Likelihood from SDE.}  An OU process is defined by the SDE
\begin{equation}
\label{eq:ou-sde}
dx_t = \mu (\theta - x_t)dt + \sigma dB_t,
\end{equation}
where $B_t$ is a standard Brownian motion under the physical probability measure. The likelihood of an OU process observed over a sequence $\{x_t\}_{t=1}^T$ is given by
\[
\begin{aligned}
\prod_{t=1}^T & f(x_t|x_{t-1}) = \prod_{t=1}^T\frac{1}{\sqrt{2\pi \tilde{\sigma}^2}} \times\\
& \exp\left(-\frac{x_t - x_{t-1}\cm - \theta (1-\cm)^2}{2\tilde{\sigma}^2}\right)
 \end{aligned}
 \] 
where $\tilde{\sigma}^2 = \sigma^2 \frac{1-\cm^2}{2\mu}$. Minimizing the negative log-likelihood results in the  optimization problem 
\begin{equation} \label{mle1}
\begin{aligned}
 \min_{\mu,\sigma^2,\theta,w} &\frac{1}{2}\ln(2\pi) + \frac{1}{2}\ln(\tilde{\sigma}^2(\mu,\sigma^2)) + \frac{ \| A(\mu)w - y(\theta,\mu)\|^2}{2T\tilde{\sigma}^2(\mu,\sigma^2)}, 
 \end{aligned}
 \end{equation}
with $y = \theta(1-\cm){\bold 1}$, and $A \in \R^{n \times 2}$ defined as 
\[ A = S_{1:T}- \cm S_{0:T-1}, \]
where the subscripts denote ranges for $t$.

\noindent
\begin{remark}
\label{rem:unbdd}
The objective function in (\ref{mle1}) is unbounded. Set $w=0, \theta = 0$; the objective function is then given by  
\[
\frac{1}{2}\ln(2\pi) + \frac{1}{2}\ln(\sigma^2) + \frac{1}{2} \ln\left(\frac{1-\exp(-2\mu\Delta t)}{2\mu}\right),
\] 
which will go to negative infinity as $\sigma^2 \to 0$. 
\end{remark}
To solve the issue exposed in Remark~\ref{rem:unbdd}, we add a 1-norm equality constraint on $w$, setting $\|w\|_1=1$.
This constraint is also convenient from a modeling perspective, as it eliminates the need to select which 
assets in the portfolio are to be long or short {\it a priori}. 

To get the formulation~\eqref{eq:mlegen}, we let 
\begin{equation}
\label{eq:varchangesde}
 a = \tilde{\sigma}^2 = \frac{\sigma^2(1-\exp(-2\Delta t \mu))}{2\mu}, c = \exp(-\Delta t \mu),
 \end{equation} 
after which the problem becomes
\[ 
\min_{a, c, \theta,\|w\|_1 = 1} \frac{1}{2} \ln(a) + \frac{1}{2Ta} \| A(c)w - \theta (1- c) \|^2.
\]
  We can recover $\mu$ and $\sigma^2$ once we know $a$ and $c$.
The change of variables~\eqref{eq:varchangesde} simplifies the model structure by encapsulating complicated parts into $a$ and $c$. It also conveniently eliminates the occurrence of $\Delta t$ in objective function. 

\noindent
{\bf Maximum Likelihood from AR.} The AR formulation arises from a simple discretization of the OU process~\eqref{eq:ou-sde}:
\begin{equation} 
\label{dou}
x_t = x_{t-1} + \mu (\theta - x_{t-1})\Delta t + \sigma \sqrt{\Delta t}\epsilon_t, \epsilon_t \sim N(0,1).
\end{equation}
With $x_t = S_tw$ being the portfolio price process, we obtain   a joint objective for AR parameters and the portfolio $w$:
\[
\min_{\sigma^2} \frac{1}{2}\ln(\Delta t\sigma^2) + \frac{1}{2T\Delta t\sigma^2}\min_{\mu,\theta,w} h(\mu,\theta,w) \]
where 
\[h(\mu,\theta,w) = \sum_{t=1}^T (S_{t-1}w +\Delta t \mu (\theta - S_{t-1}w) -S_tw)^2.\]
As before, we include the equality constraint $\|w\|_1=1$ to avoid unboundedness problem 
noted in Remark~\ref{rem:unbdd}.

To obtain formulation~\eqref{eq:mlegen}, we take 
\begin{equation}
\label{eq:varchangear}
a = \Delta t\sigma^2,\quad  c = 1- \Delta t\mu, 
\end{equation}
so that the problem becomes
\[
\min_{a ,c ,\theta,\|w\|_1=1} \frac{1}{2}\ln(a) + \frac{1}{2Ta} \| A(c)w -\theta(1- c)\|^2
\]
 where $A(c) = S_{1:T} - cS_{0:T-1}$.
Parameters $a$ and $c$ have different expressions in the SDE and AR formulations, 
with $a=\frac{\sigma^2(1-\exp(-2\Delta t \mu))}{2\mu}$ in~\eqref{eq:varchangesde} and $a=\Delta t \sigma^2$ in~\eqref{eq:varchangear}, and 
$c=\exp(-\Delta t \mu)$  in~\eqref{eq:varchangesde} and $c= 1 - \Delta t \mu$ in~\eqref{eq:varchangear}. 
The expressions converge as $\Delta t \mu \downarrow 0$. 
We use the characterizations from~\eqref{eq:varchangear} in the numerical experiments. 

\subsection{Promoting Sparsity and Mean-Reversion }
Given a set of candidate assets,  we want to select a small parsimonious subset to build a portfolio. 
To add this feature to the model, we want to impose a sparsity penalty on $w$. While the 1-norm 
is frequently used,  in our case we have already imposed the 1-norm equality constraint $\|w\|_1 = 1$. 
To obtain sparse solutions under this constraint, we add a multiple of the {\it concave} penalty 
$-\frac{\eta}{2}\|w\|^2$ to the maximum likelihood~\eqref{eq:mlegen}. The negative quadratic takes minima on the vertices of the 1-norm ball, and adding this penalty pushes the selection towards sparser $w$.  

In addition to sparsifying the solution, we may also want to promote other features of the portfolio. 
The penalized likelihood framework is flexible enough to allow these enhancements. 
An important feature is encapsulated by the mean-reverting coefficient $\mu$; 
a higher $\mu$ may be desirable. We can try for a higher $\mu$ by promoting a lower $c$, 
e.g. with a linear penalty. The augmented likelihood function is 
\begin{equation} 
\label{mle3}
  \min_{a,c, \theta,\|w\|_1 = 1}  \frac{\ln(a)}{2}   + \frac{\| A(c)w - \theta (1-c)\|^2}{2T a}  + \gamma c - \frac{\eta}{2}\|w\|_2^2.
  \end{equation}
  
\section{Algorithm}
\label{sec:algo}

In this section we develop an algorithm to solve the \emph{nonsmooth},  \emph{nonconvex} problem~\eqref{mle3} by exploiting 
its rich structure. 
We define the following nested value functions:
\begin{eqnarray}
f(w, a, c, \theta) &=& \displaystyle- \frac{\eta\|w\|_2^2}{2} + \frac{\ln(a)}{2} + \gamma c +\frac{\| A(c)w - \theta (1-c)\|^2}{2T a}\nonumber\\
f_1(w, a, c) &=& \min_\theta f(w, a, c, \theta) \nonumber \\
f_2(w, a) &=& \min_c f_1(w,a,c) = \min_{c,\theta} f(w, a, c, \theta)\nonumber \\
f_3(w) & =& \min_{a} f_2(w, a) =\min_{a, c, \theta} f(w, a, c, \theta). \label{eq:vfs}
\end{eqnarray}
 Our main strategy is to recast~\eqref{mle3} as an optimization problem 
\[
\min_{\|w\|_1 = c}f_3(w),
\]
where every evaluation of $f_3$
is done by partially minimizing over the remaining parameters,
efficiently implemented using  $f_2$ and $f_1$. 
Partial minimization strategies can significantly accelerate 
solutions of complex structured problems~\cite{aravkin2012estimating,aravkin2017efficient}.

We start with the $f_1$ subproblem. Taking $\partial_\theta f  = 0$, we get  
\[ 
\begin{aligned}
0 = \frac{\partial f}{\partial \theta}  &= (1-c){\bf 1}^T(\theta ( 1-c) - A(c)w ) \\
\Rightarrow \theta^*& = \frac{{\bf 1}^T(x(w)_{1:T} - c x(w)_{0:T-1})}{T(1-c)}.
\end{aligned}
\]
Plugging $\theta^*(c,w)$ into $f$, we get an explicit form of $f_1$:
\begin{align}
f_1(w,a,c) &= - \frac{\eta}{2}\|w\|_2^2 + \frac{1}{2} \ln(a)  + \gamma c \notag\\
&+ \frac{1}{2T a}\| B(x(w)_{1:T} - cx(w)_{0:T-1})\|^2  
\end{align}
with  $B = {\bf I} - \frac{{\bf 1}{\bf 1}^T}{T}$ a projection matrix onto the space of vectors in $\mathbb{R}^T$ with mean $0$.

We now minimize with respect to $c$. 
\[  
\begin{aligned}
0 = \frac{\partial f_1}{\partial c}  &= \gamma + \frac{1}{Ta}x_{0:T-1}^TB^TB(c x(w)_{0:T-1} - x(w)_{1:T}) \\
  \Rightarrow c^*(a,w)  & = \frac{(Bx(w)_{0:T-1})^T(Bx(w)_{1:T}) - Ta\gamma}{\|Bx(w)_{0:T-1}\|^2}.
  \end{aligned}
   \]
We can use this expression to explicitly write $f_2(w,a)$:
\begin{equation} \label{aw}
\begin{aligned}
 f_2(a,w) & = - \frac{\eta}{2}\|w\|_2^2 + \frac{1}{2} \ln(a)  + \gamma c^*(a,w) \\
 &+ \frac{1}{2T a}\| B(x(w)_{1:T} - c^*(a,w)x(w)_{0:T-1})\|^2.
 \end{aligned}
  \end{equation}

\begin{figure}[h!]
\centering
\includegraphics[scale = .6]{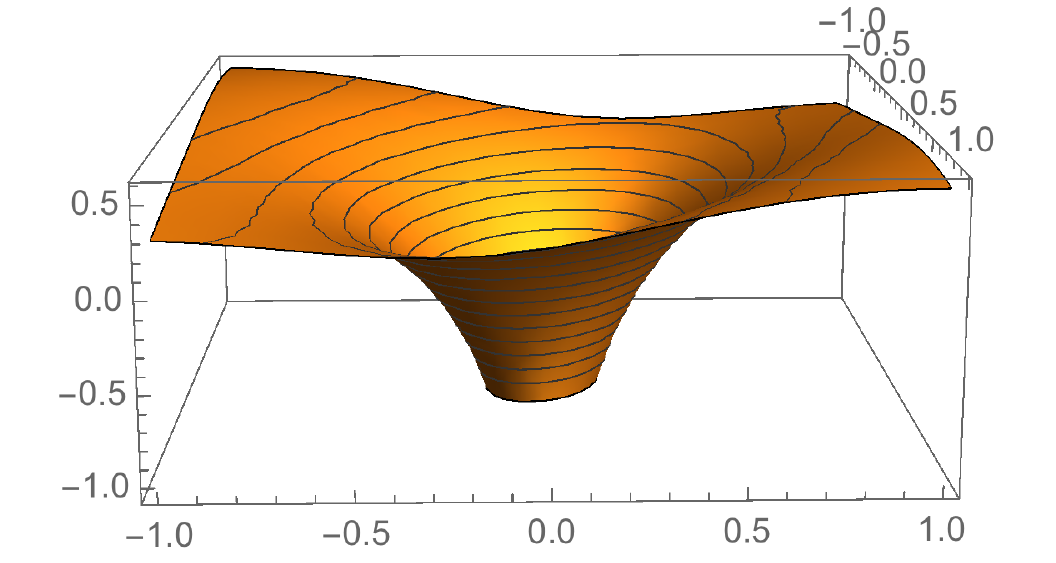}
\caption{\label{fig:concvf}3D plot for objective function in~\eqref{eq:ccvgam0}, for $w\in\mathbb{R}^2$.}
\end{figure}

{\bf Structure of objective when $\gamma = 0.$}
We first consider the case $\gamma = 0$. 
In this case, $c^*(a,w) = c^*(w)$ is independent of $a$. 
We can then minimize $f_2$ in $a$:
\begin{equation} 
\label{var}
\begin{aligned}
0 = \frac{\partial f_2}{\partial a} & = \frac{1}{2a}  - \frac{1}{2Ta^2} \| B(x_{1:T} - c^*x_{0:T-1})\|^2 \\
\Rightarrow a^*(w)& = \frac{\| B(x(w)_{1:T} - c^*(w)x(w)_{0:T-1})\|^2}{T}. 
\end{aligned}
\end{equation}
Plugging $a^*$ in, we get a closed-form expression for $f_1$: 
\[
\begin{aligned}
f_1(w)  &=  - \frac{\eta}{2}\|w\|_2^2 + \frac{1}{2}\ln(\| B(x_{1:T} - c^*x_{0:T-1})\|^2) \\
& = - \frac{\eta}{2}\|w\|_2^2 + \frac{1}{2}\ln(\| BA(c^*(w))w\|^2)
\end{aligned}
\]
The optimization problem in this case reduces to 
\begin{equation}
\label{eq:ccvgam0}
\min_{\|w\|_1 = 1} - \frac{\eta}{2}\|w\|_2^2 + \frac{1}{2}\ln(\| BA(c^*(w))w\|^2),
\end{equation}
which we can solve with projected gradient descent. The set $\|w\|_1 = 1$ is nonconvex, but has an easy projection:
\[\begin{aligned}
w & \gets \text{argmin}_{\|z\|_1=1}\|w - z\|^2 \\
& = \text{argmin}_{\|z\|_1=1} \| |w| - \frac{1}{\text{sign}(w)} \odot z\|^2 \\
& = \text{sign}(w) \odot \text{argmin}_{\|u\|_1=1} \| |w| - u \|^2 \\
& =  \text{sign}(w)\odot  \text{argmin}_{\|u\|_1=1,u \geq 0 } \| |w| - u \|^2\\
& =  \text{sign}(w) \odot \text{proj}_{\Delta_1}(|w|), \quad \Delta_1 \; \mbox{ the unit simplex},
\end{aligned}
\]
where $\odot$ is elementwise multiplication, and  
the second equality is obtained by a change of variable $u =  \frac{1}{\text{sign}(w)} \odot z$. 
Fast projectors onto the unit simplex $\Delta_1$ are well-known~\cite{duchi2008efficient}. 
\begin{remark}
\label{rem:even}
The objective function of~\eqref{eq:ccvgam0} is an even function. Hence the problem does not have a unique minimizer. See also 
Figure \ref{fig:concvf} for a 3D plot of this function when $w\in\mathbb{R}^2$.
\end{remark}

\begin{remark}
\label{rem:nonconvex}
The objective function of~\eqref{eq:ccvgam0}  is nonconvex, and its Hessian is indefinite. 
The Hessian is given by 
\[ 
H = -\eta I + \frac{1}{\|Dw\|^2} D^T\left(I -2 \left (\frac{Dw}{\|Dw\|}\right)\left (\frac{Dw}{\|Dw\|}\right)^T\right) D 
\]
where $D = BA(c^*)$. The matrix  $I - 2\left (\frac{Dw}{\|Dw\|}\right)\left (\frac{Dw}{\|Dw\|}\right)^T$ is a 
Householder reflection matrix, with one eigenvalue equal to -1 and the rest 1.
\end{remark}
\begin{algorithm}
  \caption[Caption]{\label{pgmle}Projected Gradient Descent for $f_3(a,w)$~\eqref{eq:vfs}.} 
  \begin{algorithmic}[1]
    \Require{$w \in \R^m, S, a \in (0,\infty),f, \gamma,\eta, \epsilon$} \\
    $B =  {\bf I} - \frac{{\bf 1}{\bf 1}^T}{T}$
    \For{$i = 1,2,3,... $} 
       \Let {$x$}{$Sw$}
    	\Let{$c$}{ $\frac{(Bx_{0:T-1})^T(Bx_{1:T}) - Ta\gamma}{\|Bx_{0:T-1}\|^2}$}
	 \Let{$A$}{$x_{1:T} - cx_{0:T-1}$}
	\If{$\gamma = 0$}
		\Let{$a$}{$\|BAw\|^2/T$}
	\Else
	 \Let{$a$}{$\text{Proj}_{(0,\infty)}\left(a - \delta_{a,i}\partial_a f (w,a)\right)$}
	 \EndIf
      	\Let{$w$}{$\text{Proj}_{\|\cdot\|_1 = 1}\left(w - \delta_{w,i}\partial_w f (w,a)\right)$} 
	\Let{loss$_{i}$}{$f(w,a)$} \\
	Iterate till convergence in loss
      \EndFor
  \end{algorithmic}
\end{algorithm}
{\bf Structure of the objective when $\gamma >0$.} When $\gamma > 0$, $c^*$ depends on $a$. 
Plugging $c^*(a,w)$ into $f_2$ in~\eqref{aw}, and taking  
\[
b_1(w) = Bx(w)_{1:T}, \quad b_0(w) = Bx(w)_{0:T-1},
\]
we can write the value function $f_3(w)$ as 
\begin{align*}
\min_a  
& \frac{1}{2}\ln(a) +  \frac{\|b_1(w)\|^2}{2Ta} - \frac{ (b_0(w)^Tb_1(w) - Ta\gamma)^2}{2Ta \|b_0(w)\|^2},
\end{align*}
which is equivalent to 
\begin{equation} \label{agamma}
 \min_a \frac{1}{2}\ln(a) + \frac{\|b_1\|^2}{2Ta} - \frac{(b_0^Tb_1)^2}{2Ta \|b_0\|^2} - \frac{Ta\gamma^2}{2\|b_0\|^2}. 
 \end{equation}
Solving \eqref{agamma}, we obtain the optimal $a$ in  closed form:
 \[ a^* = \frac{\|b_0\|^2}{2T\gamma^2} - \frac{\sqrt{\|b_0\|^4 - 4\gamma^2(\|b_0\|^2\|b_1\|^2 - (b_0^Tb_1)^2)}}{2T\gamma^2}\]
and correspondingly 
 \[
  c^* = \frac{b_0^Tb_1}{\|b_0\|^2} - \frac{1}{2\gamma} + \sqrt{ \frac{(b_0^Tb_1)^2}{\|b_0\|^4} + \frac{1}{4\gamma^2} - \frac{\|b_1\|^2}{\|b_0\|^2}}.
  \]
\begin{figure}
\center
   \includegraphics[scale = .23]{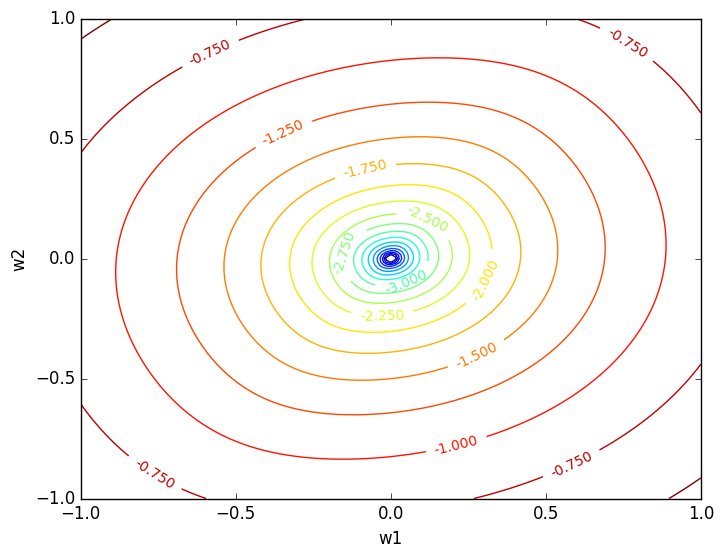}\includegraphics[scale = .23]{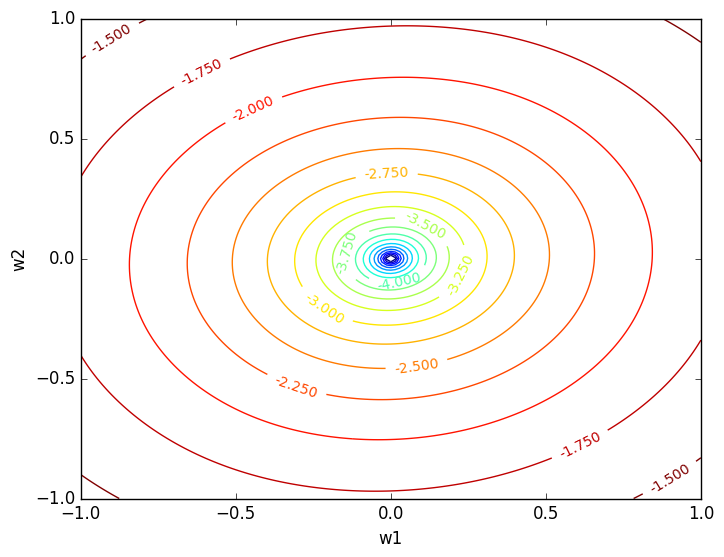}
   \caption{\label{g0}Contour plot of~\eqref{aw} $\gamma \neq 0$ (left) and $\gamma = 0$ (right) for $w\in\mathbb{R}^2$.}
\end{figure}
 In these expressions, $b_0$ and $b_1$ are functions of $w$. The optimal solution 
 $a^*$ increases with respect to $\gamma$ and $c^*$ decreases with respect to $\gamma$ when 
 \[ 0 < \gamma <  \frac{1}{2}\sqrt{\frac{\|b_0\|^4}{-(b_1^Tb_0)^2 + \|b_0\|^2\|b_1\|^2}}. \]

In this general case, we can also write down the final optimization problem in closed form: 
\[
f_1(w) =  \frac{1}{2}\ln(a^*) + \frac{\|b_1\|^2}{2Ta^*} - \frac{(b_0^Tb_1)^2}{2Ta^* \|b_0\|^2} - \frac{Ta\gamma^2}{2\|b_0\|^2},
\]
where $b_i$ and $a^*$ are all functions of $w$ as detailed above.  
Figure \ref{g0} illustrates the effect of $\gamma$ on the shape of contours for a simple case where $w$ has dimension 2. 
The contours are rotated by $\gamma$, which can affect which assets are selected (as it can change the intersection 
points with the 1-norm ball). 

In the general case $\gamma > 0$, we work with $f_2(a,w)$ rather than with $f_3(w)$, 
to avoid dealing with the highly nonlinear form of $a(w)$. The final approach is detailed in Algorithm~\ref{pgmle}.


\section{Numerical Results}
\label{sec:numerics}

\subsection{Single Time Series}
We  demonstrate that given an observed time series from an OU process, 
our formulation can recover true underlying OU parameters. 
Given a portfolio $w$, we   solve for optimal parameters from 
\[ 
\min_{a,c,\theta} \frac{1}{2}\ln(a) + \frac{1}{2Ta}\| x_{1:T} - cx_{0:T-1} -\theta(1-c)\|^2.
\]
We obtain multiple realizations of time series using the discretized OU~\eqref{dou}, 
estimated parameters from the realizations, and compute the average deviation of our estimate 
from the true generating parameters. 
We also study the effect of varying $\Delta t$ and total time span $L$ on the estimation. 
The number of total time points is given by  $T = L/\Delta t$.

\begin{figure}[h]
   \centering
   \begin{tabular}{c} 
   \includegraphics[scale = .5]{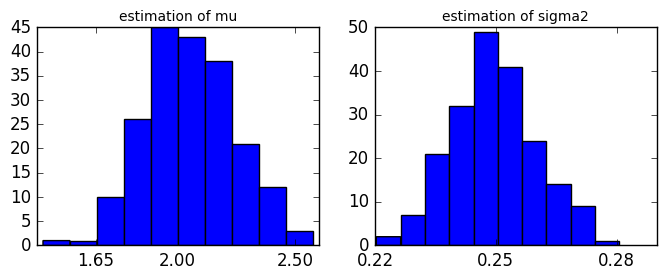} \\
   \includegraphics[scale = .5]{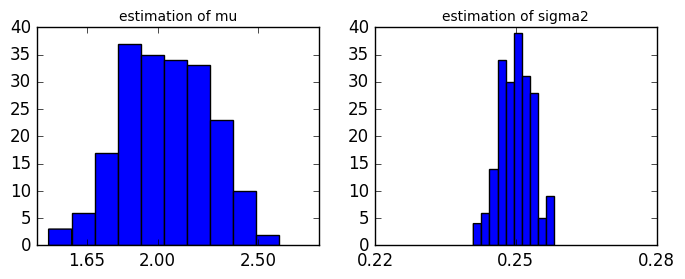} \\
   \includegraphics[scale = .5]{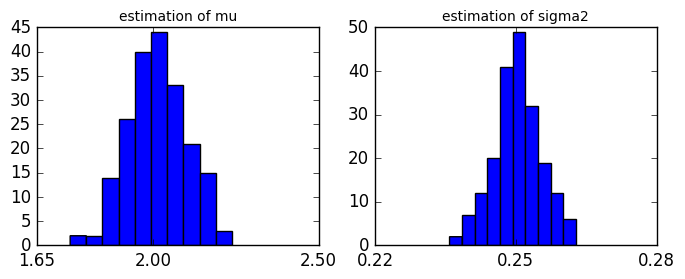}
   \end{tabular}
   \caption{ \label{sg}
\small Distribution of estimated $\mu$ and $\sigma^2$. $\mu_{true} = 2, \sigma_{true}^2 = 0.25$. Top panel: $\Delta t = 0.1, L = 100$; mid panel: $\Delta t = 0.01, L = 100$; bottom panel: $\Delta t = 0.1, L = 500$.}
\end{figure}

Figure \ref{sg} shows results from one experiment with true parameters $\mu = 2$ and $\sigma^2 = 0.25$. Top row of the figure corresponds to $\Delta t = 0.1, L = 100$. For the middle row, we decrease $\Delta t$ to 0.01 but keep $L$ unchanged. The deviation of estimates of $\sigma^2$ from the true parameter decreased significantly, while deviation of estimates of $\mu$ from the true parameter remained the same. 
In the bottom row, we keep $\Delta t = 0.1$ but increase $L$ to $500$. 
Estimation of both parameters improved significantly. 
The average deviation of estimated parameters, defined as
$\frac{| \mu - \hat\mu|}{|\mu|}$ and $\frac{| \sigma^2 - \hat \sigma^2|}{|\sigma^2|}$, 
are summarized in Table \ref{sgd}.

\begin{table}[h]
   \centering
   \begin{tabular}{cc|cc} 
   \hline
   $\Delta t$ & $L$ & dev($\mu$) & dev($\sigma^2$) \\ \hline
    0.1 & 100 & 0.07 & 0.03\\
    0.01 & 100 & 0.08 & 0.01 \\
    0.1 & 500 & 0.03 & 0.01 \\
    \hline
   \end{tabular}
   \caption{\small Average deviation of $\hat \mu$ and $\hat \sigma^2$ from $\mu = 2, \sigma^2 = 0.25$. }
   \label{sgd}
\end{table}
\subsection{Selection for Multiple Time Series}

\begin{figure}
   \begin{tabular}{c}
   \includegraphics[scale = .38]{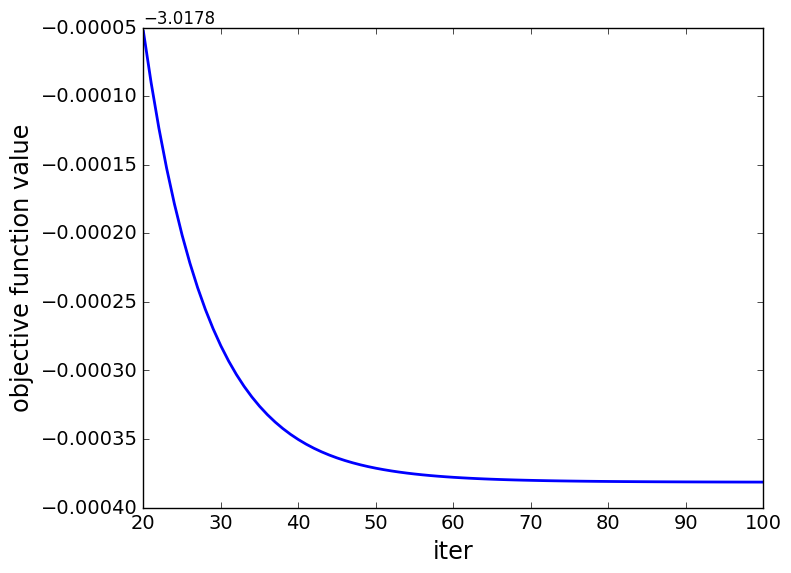} \\
      \includegraphics[scale = .4]{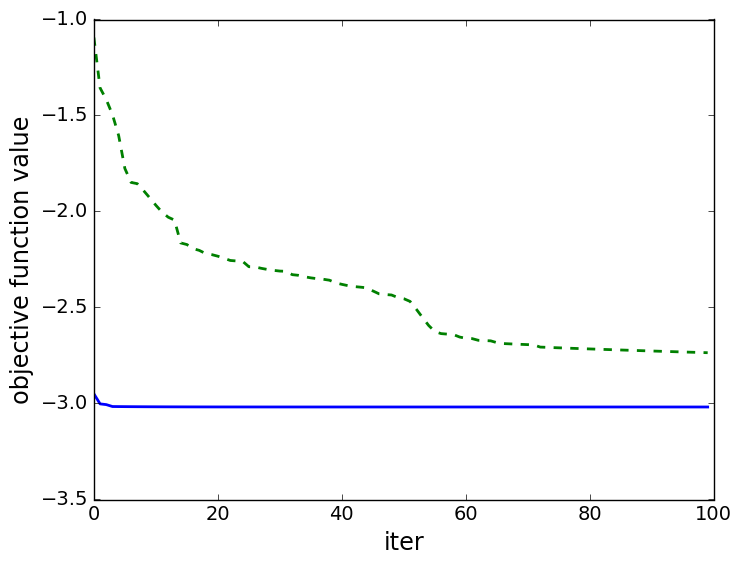}
      \end{tabular}
   \caption{\label{cvgsim} Top panel: zoom-in on objective decrease for Algorithm~\ref{pgmle}. Bottom panel: comparison of Algorithm~\ref{pgmle} (solid) with standard projected gradient (dashed) using objective function values.}
   \end{figure}

{\bf Algorithmic Comparison using Simulated Data}. In our first experiment, we show (1) 
that we can identify mean-reverting time series using simulated data and (2) that Algorithm~\ref{pgmle} 
is faster than a standard approach that does not use partial minimization. 
We simulate five time series; four from an OU process with different $\mu$ and $\sigma$ as specified in Table \ref{sim}, and one is non-OU time series with $\sigma = .1$. 
All have $T = 500$ and $\Delta t = 0.01$. We use the first 70\% of data for training and 30\% for testing. Figure \ref{cvgsim} shows convergence plots in objective function values. Top panel shows the plots using Algorithm~\ref{pgmle}, and bottom panel shows the comparison between that and regular projected gradient descent on all unknowns (without partial minimization). 

  \begin{table}[h!]
 \centering
 \begin{tabular}{c|cccccc}\hline
 \# & 1 & 2 & 3 & 4  \\
  \hline
   $\mu$ & 1 & 4 & 1 & 4 \\
   $\sigma$ & 1 & 1 & 0.5 & 0.5 \\
   $\theta$ & 0 & 1 & 1 & 0 \\
   \hline
   \end{tabular}
   \caption{\label{sim}Model parameters for simulated OU time series}
   \end{table}
   
\begin{figure}[h!]
   \centering
   \includegraphics[scale = .45]{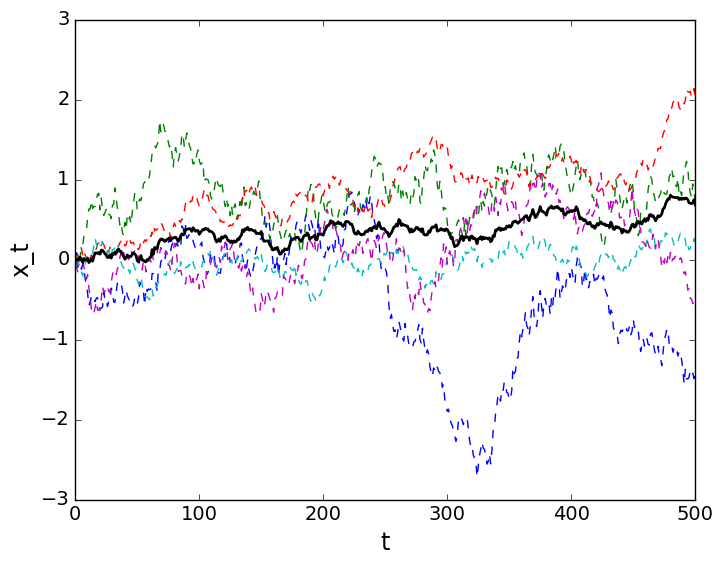}
   \caption{\label{sim_plot}Plot of simulated asset time series (dashed) and time series for final selected portfolio (solid).}
   \end{figure}

When $\gamma = \eta = 0$, the estimated parameter values are $\mu = 2.86, \sigma^2 = 0.09, \theta = 0.31$ with weight vector $w = [0.10, 0.10,0.36,0.33,0.11] $. The model puts 69\% of the weights into the pair of OU time series with $\sigma = 0.5$. 
In other words, it favors OU time series with a  lower $\sigma$ value  but remains relatively indifferent to $\mu$ values. Figure \ref{sim_plot} plots those time series and the portfolio selected by the model.

Table \ref{tune1} compares process parameters and weight vectors as we tune $\gamma$ and $\eta$.
In particular, by increasing $\eta$ to 1.5, we obtain a portfolio which places $80\%$ rather than $69\%$ of the weights into the 
two assets. Increasing $\gamma$ to 0.05 gives a very small preference to the time series with a slightly larger $\mu$. When both $\gamma$ and $\eta$ are tuned, as shown in the last row of the table, the model picks the fourth time series, the one with larger $\mu$ and smaller $\sigma$ compared to other OU time series.
  \begin{table}[h!]
 \centering
 
 \begin{tabular}{c|c|c|c|c|c}\hline
 $\gamma$ & $\eta$ &$\mu$ & $\sigma^2$ & $\theta$ & $w$  \\
  \hline
   0 & 0 & 2.86 & 0.09 & 0.31 &  [0.10, 0.10,0.36,0.33,0.11] \\
   0.05 & 0 & 3.13 & 0.09 & 0.30 & [0.10, 0.10,0.36,0.33,0.11] \\
   0 & 1.5 & 2.35 & 0.09 & 0.36 & [0.07, 0.06,0.44,0.36,0.07]  \\
   0.05 & 1.5 & 5.53 & 0.27 & -0.06 & [0.0, -0.0, 0.0, 1.0, 0.0]\\ \hline

   \end{tabular}   \caption{\label{tune1}Model Estimations with $\gamma$ and $\eta$}
   \end{table}

{\bf Real data.}
We performed experiments with empirical price data from three groups of selected assets: precious metals, large   equities and oil companies, see Table \ref{ag}. Data were taken from Yahoo Finance, and give closing stock prices for each asset over the past five years. The first 70\% of data (over time) is used for training, and the rest for testing. 

For each group, we progressively augmented the set of candidate assets, and applied our approach.  
The negative log-likelihoods of portfolios selected from each set are in Table \ref{pm}, 
along with negative log-likelihoods of individual assets. 
For the precious metals group, we considered two different orderings of growing candidate asset sets.  

Table \ref{pm} shows that portfolio negative log likelihoods are generally smaller than negative log likelihoods of individual assets and decrease as we include more assets, which means we can obtain more OU-representable portfolios as the candidate sets expand. The relative improvement depends on the assets, as well as the ordering. 
For instance, for precious metals we observed a plateau as we add GG to the universe with the first ordering, 
but not with the second. These  results are reminiscent of those for step-wise model selection. Our approach can take a group of multiple candidate assets and find an OU-representable portfolio from a subset, and we can discover diminishing return from further expanding the groups of candidate assets.

  \begin{table}[h]
 \centering
 \begin{tabular}{c|c}\hline
 Groups & Assets (Tickers) \\
  \hline
   precious metals & GLD GDX, GDXJ, SLV, GG, ABX \\
   large equities & GOOG, JNJ, NKE, MCD, SBUX, SPY, VIG, VO \\
   oil companies & BP, COP, CVX, OIL, USO, VLO, XOM\\
   \hline
   \end{tabular}
    \caption{\label{ag}Asset Groups for Empirical Experiments}
   \end{table}

\begin{figure*}
 \center
 \begin{tabular}{cc}
  (a) precious metals & \\
 \includegraphics[scale = .4]{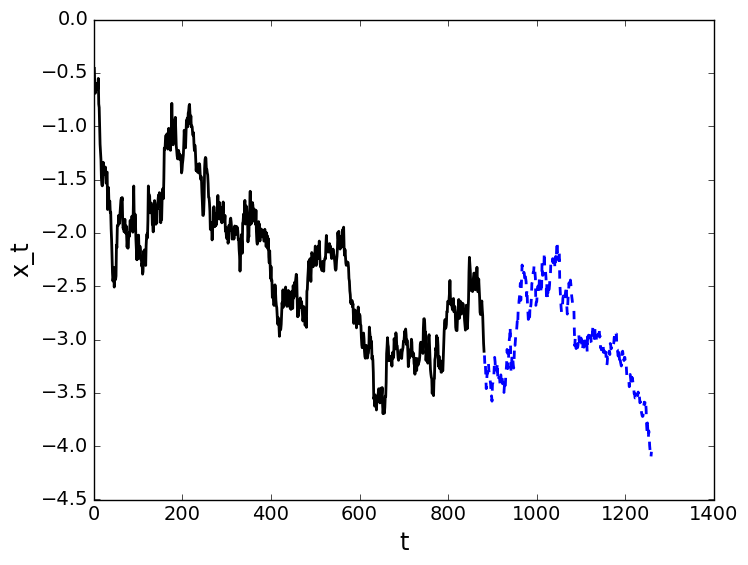} &
   \includegraphics[scale = .4]{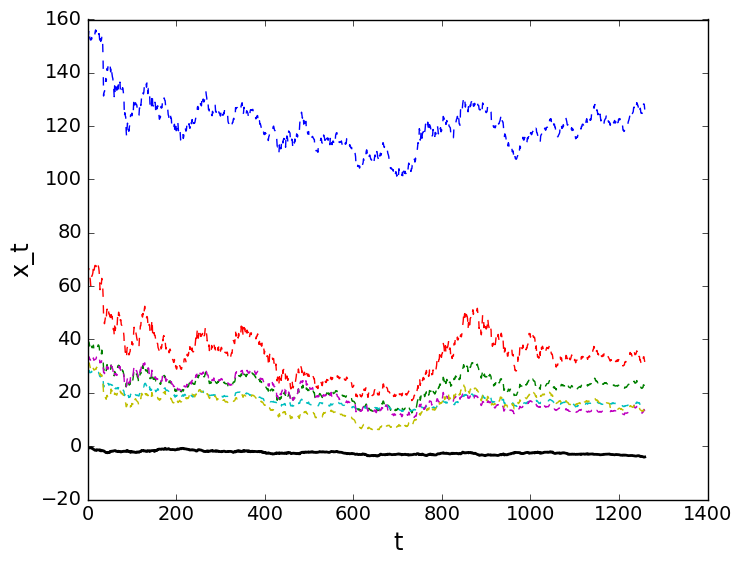} \\
   (b) large capital equities & \\
  \includegraphics[scale = .41]{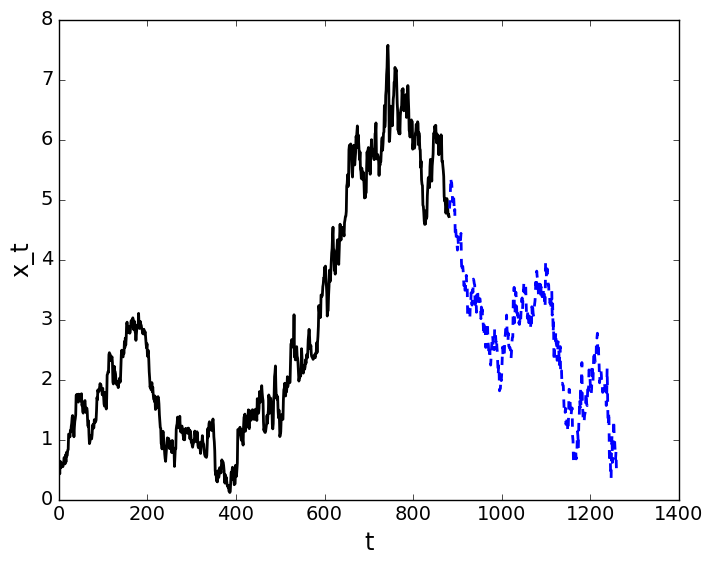} &
    \includegraphics[scale = .4]{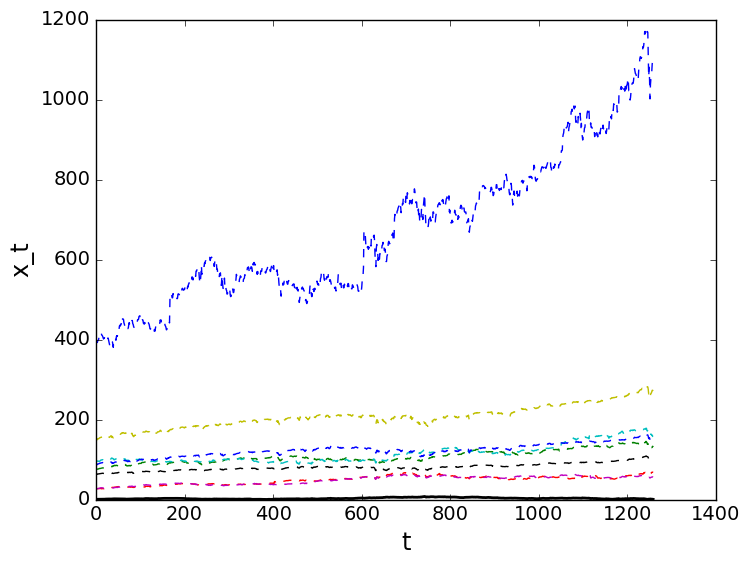} \\
    (c) oil companies &\\
   \includegraphics[scale = .4]{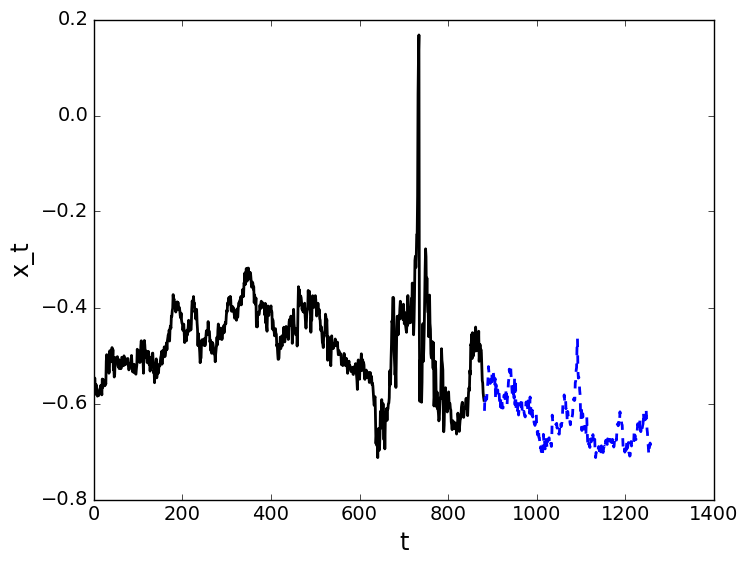} &
     \includegraphics[scale = .4]{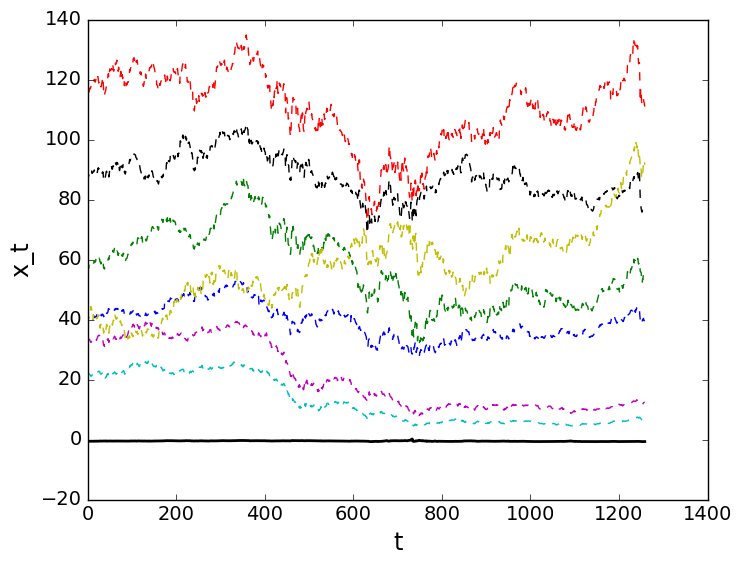}
   \end{tabular}
   \caption{\label{port} Model selected portfolio time series from real data.}
   \end{figure*}

In  Figure \ref{port} we plot the model selected portfolio for each group (left), and compare it to assets within its group (right). As seen, our model constructs a portfolio that mean-reverts even though some of the assets show significant  trends.

 
 We also conducted experiments including $\gamma$ and $\eta$, to promote larger $\mu$ and sparser portfolios. 
 The results are summarized in Table \ref{tune1}. 
 When $\gamma > 0$, we see increasing $\mu$ across asset groups. 
 As we increase $\eta$, for precious metals we see more concentrated weights on SLV: 41\% when $\eta = 0$, 50\% when $\eta = 2.7$ and $84\%$ when $\eta = 3, \gamma = 0.2$. For large capital equities, we see a significant shift when $\eta = 3.5,\gamma = 0.1$, getting a majority of the weight on VIG. For oil companies the selection was already sparse so there is only a 
 slight change in weight when we increase $\eta$.    
 
When $\eta$ is sufficiently large, it pushes the model to select only one asset from a given group, 
i.e. $w_i \approx 1$ for some $i$ and 0 elsewhere. 
Ideally we would want the selected asset to correspond to the one having minimum negative log-likelihood
across the asset class. However, as the problem is nonsmooth and nonconvex, the selected index
 $i$ can   be sensitive to the initialization of $w$ for large $\eta$. 
 
 \begin{table}
 \centering
 \begin{tabular}{l|c|l|l}\hline
 Index $i$ & Assets & Portfolio with assets up to $i$ & Individual\\
  & & (train, test) & (train, test) \\
 \hline
 1 & GLD & & 0.77, 0.44\\
 2 & GDX & -0.77, -1.02 & 0.05, -0.30\\
 3 & GDXJ &-1.48,-1.77 & 0.70, 0.38\\
 4 & SLV & -1.86, -2.10 & -0.69, -1.0\\
 5 & GG & -1.86, -2.10 & -0.04, -0.44 \\
 6 & ABX & -1.89, -2.10& -0.24 ,-0.54\\
 \hline \hline
 Index $i$ & Assets & Portfolio with assets up to $i$ & Individual\\
  & & (train, test) & (train, test) \\
 \hline
 1 & GLD & & 0.77, 0.44\\
 2 & GDXJ & 0.66, 0.46 & 0.05, -0.30\\
 3 & SLV &-1.49,-1.72 & 0.70, 0.38\\
 4 & GG& -1.67, -1.89 & -0.69, -1.0\\
 5 & GDX & -1.86, -2.11 & -0.04, -0.44 \\
 6 & ABX & -1.89, -2.10& -0.24 ,-0.54\\
  \hline \hline
 Index $i$ & Assets & Portfolio with assets up to $i$ & Individual\\
  & & (train, test) & (train, test) \\
 \hline
 1 & GOOG & & 2.66, 3.06\\
 2 & JNJ & 0.27, 0.64 & 0.40, 0.86\\
 3 & NKE &-0.13,0.12 & 0.09, 0.43\\
 4 & MCD & -0.15, 0.04 & 0.49, 1.09\\
 5 & SBUX & -0.29,-0.20 & 0.02, 0.05 \\
 6 & SPY & -0.84, -0.55& 0.95 ,1.00\\
 7 & VIG & -1.26, -0.89& -0.07 ,0.01\\
  8 & VO & -1.42, -1.05& 0.53 ,0.45\\
   \hline \hline
 Index $i$ & Assets & Portfolio with assets up to $i$ & Individual\\
  & & (train, test) & (train, test) \\
 \hline
 1 & BP & & -0.09, -0.33\\
 2 & COP & -0.75, -0.86 & 0.46, 0.25\\
 3 & CVX &-0.77,-0.77 & 0.79, 0.73\\
 4 & OIL & -1.16, -1.33 & -0.84, -1.25\\
 5 & USO & -2.91,-3.26 & -0.45, -0.86 \\
 6 & VLO & -2.89, -3.28& 0.58, 0.43\\
 7 & XOM & -2.95, -3.31& 0.48, 0.26\\
 \hline
   \end{tabular}
   \caption{\label{pm}Negative log-likelihood of assets groups}
   \end{table}
   
  \begin{table}[h!]
 \centering
 \begin{tabular}{cc|ccc|c}\hline
 {Group}&1 & & & & \\
 \hline
 $\gamma$ & $\eta$ &$\mu$ & $\sigma^2$ & $\theta$ & $w$  \\
  \hline
   0 & 0 & \textcolor{blue}{3.14} & 2.11 & -2.53 &  [-.09,.18,-.15,\textcolor{red}{.41},.02,.14] \\
   0.2 & 0 & \textcolor{blue}{4.39} & 2.11 & -2.42 &  [-.09, .19, -.15,0.4,0.02,0.13] \\
   0 & 2.7 & 3.92 & 1.95 & -2.48 & [-.09,.13,-.08, \textcolor{red}{.50},-.08,.1]  \\
   0.2 & 3 & 3.13 & 4.75 & -3.58 & [-.15, 0, -.01, \textcolor{red}{0.84}, 0, 0] \\
   \hline\hline
Group &2 & & & & \\
 \hline
 $\gamma$ & $\eta$ &$\mu$ & $\sigma^2$ & $\theta$ & $w$  \\
  \hline
   0 & 0 & \textcolor{blue}{.72} & 5.44 & 4.49 &  [0,.08,0.13,.02,.16, \textcolor{red}{-.26}, \textcolor{red}{.22},.12] \\
   0.1 & 0 & \textcolor{blue}{0.86} & 5.46 & 4.27 &  [0, .08, .13,.02,.16,-.26,.22,.12] \\
   0 & 3.2 & .6 & 5.78 & 4.96 & [0,.08,.13, .01,.18, -.26, .24,.1]  \\
   0.1 & 3.5 & 2.06 & 3 & 6.13 & [0, -.01, 0, -.01, 0, \textcolor{red}{-.25}, \textcolor{red}{0.73}, 0] \\
   \hline\hline
Group &3 & & & & \\
 \hline
 $\gamma$ & $\eta$ &$\mu$ & $\sigma^2$ & $\theta$ & $w$  \\
  \hline
   0 & 0 & \textcolor{blue}{11.49} & .25 & -.34 &  [.02,.01,-.01, .55, -.4,0,.01] \\
   0.1 & 0 & \textcolor{blue}{22.75} & .25 & -.66 &   [.02,.01,0,.56,-.41,0,-.01] \\
   0 & 2.7 & 18.42 & .25 & -.52 & [.01,.01,0, .57, -.41,0,0]  \\
   0.1 & 3 & 13.37 & .25 & -.40 & [.02, .01, -.01, .55, -.40, 0, .01] \\ \hline
   \end{tabular}
   \caption{\label{tune1}Model Estimations with Different $\gamma$ and $\eta$ for asset groups}
   \end{table}
   

\section{Discussion} In this paper, we solved  a joint optimization problem for simultaneous portfolio selection and OU-fitting, incorporating the quality of OU representation into the portfolio construction. We also extended 
the formulation to incorporate desirable portfolio features, including higher mean-reversion and
sparser portfolios, both important for practical trading purposes. We developed a fast algorithm for the nonsmooth nonconvex optimization problem, and presented our solutions using both simulated and real data, resulting in useful portfolios from several asset classes. Our study can motivate more sophisticated modeling to include e.g.  price dependency, dynamic portfolios, and trading decisions. 

\label{sec:discussion}
%

%

\bibliographystyle{IEEEtran}
	\bibliography{mybib}

\end{document}